\documentclass[twocolumn, superscriptaddress, aps, pra]{revtex4-2}
\usepackage{enumitem}
\usepackage[utf8]{inputenc}
\usepackage[colorinlistoftodos, color=green!40, prependcaption]{todonotes}
\usepackage{multirow}
\usepackage[normalem]{ulem}
\usepackage{amsmath}
\usepackage{url}
\usepackage{tabularx}

\usepackage{siunitx}
\usepackage[pdftex, pdftitle={Article}, pdfauthor={Author}]{hyperref}
\usepackage{booktabs}

\usepackage{graphicx}
\usepackage{dcolumn}
\usepackage{bm}
\newcommand{\filling}{\mathcal{F}_r}
\usepackage{lineno}
\usepackage{soul}

\begin{document}

\title{Controlling the centre of mass motion of levitated particles using structured wavefronts}

\author{Shah Jee Rahman}
\affiliation{Centro de Fisica de Materiales (CFM), CSIC-UPV/EHU}
\affiliation{Donostia International Physics center (DIPC)}

\author{Quimey Pears Stefano}
\email{quimeymartin.pearsstefano@ehu.eus}
\affiliation{Centro de Fisica de Materiales (CFM), CSIC-UPV/EHU}

\author{Angel Cifuentes}
\affiliation{LORTEK, Basque Research and Technology Alliance (BRTA), Arronamendia kalea 5A, 20240, Ordizia, Spain}

\author{Jason T Francis}
\affiliation{Centro de Fisica de Materiales (CFM), CSIC-UPV/EHU}
\affiliation{Donostia International Physics center (DIPC)}

\author{Iker Gómez-Viloria}
\affiliation{Centro de Fisica de Materiales (CFM), CSIC-UPV/EHU}
\affiliation{University of the Basque Country UPV/EHU}

\author{Rub\'en Pellicer‐Guridi}
\affiliation{Centro de Fisica de Materiales (CFM), CSIC-UPV/EHU}

\author{Miguel Varga}
\affiliation{Centro de Fisica de Materiales (CFM), CSIC-UPV/EHU}

\author{Gabriel Molina-Terriza}
\email{gabriel.molina.terriza@gmail.com}
\affiliation{Centro de Fisica de Materiales (CFM), CSIC-UPV/EHU}
\affiliation{Donostia International Physics center (DIPC)}
\affiliation{Ikerbasque foundation, Basque Country}

\begin{abstract} 
Optically levitated particles have great potential to form the basis of novel
quantum-enhanced sensors. These systems are very well suited for inertial
sensing, as the particles are isolated from the environment when they are
levitated at low pressures. However, there are many challenges in the
experimental realization that may affect the performance of these systems. For
example, optical aberrations in the wavefront of the trapping laser which arise
from optical elements or misalignment have a great impact on the trapping
potential. The detrimental effect of optical aberrations has not been thoroughly
studied, and usually they are iteratively corrected, giving some conflicting
results depending on the figures of merit that are used. In this work, we
present a thorough study of the effects of structuring the wavefront of the
trapping beams. We observe that clean beams, i.e. highly focused beams with
unaberrated wavefronts, may be used to optimize the longitudinal frequencies, at
the cost of the transversal ones. Our work is based in a combination of
experimental studies using a complete basis of orthogonal polynomials (Zernike
polynomials) to control the wavefront and a set of numerical calculations, which
allow us to compare the impact of structured wavefronts on the quality of traps
for optically levitated particles in vacuum. This will have direct applications
in quantum sensing and fundamental studies of quantum mechanics, as it allows
the reduction of optical backaction and thermal decoherence of the particles.

\end{abstract}

\date{\today}
\keywords{}

\maketitle

\section{Introduction}

Optical levitation of micro and nanoparticles by the radiation pressure of the
light field has become an important platform for researchers since Ashkin’s
groundbreaking experiments on optical tweezers
\cite{PhysRevLett.24.156,doi:10.1063/1.88748}. Nanoparticles levitated in vacuum
are very well isolated from the room temperature environment and have a very
high Q-factor \cite{PhysRevLett.132.133602}, these systems have demonstrated
force sensing up to $10^{-21}$N \cite{LIANG202357}. Therefore, they have been
used for weak force detection \cite{PhysRevLett.105.101101}, precise measurement
of electric and magnetic field sensing \cite{hoang2016electron,Zhu:23}, and
fundamental physics studies \cite{Aspelmeyer14}. Recently, it has been shown
that a particle trapped in a linearly polarized beam can be used as a torque
detector \cite{Ahn20}. Transferring of angular momentum to the particle in a
circularly polarized beam can induce rotations at speeds of GHz
\cite{Jin2021,PhysRevLett.121.033602,Ahn20}, providing insight into, e.g., the
vacuum friction \cite{PhysRevLett.105.113601}. Moreover, the two-dimensional
center-of-mass (CoM) motion of a single silica nanoparticle has been cooled down
to the motional quantum ground state \cite{Deli892, piotrowski2023simultaneous},
and in the near future quantum superposition of non-Gaussian states can be
realized \cite{PhysRevLett.132.023601}. This system is a very good candidate for
falsifying the wave-function collapse model \cite{RevModPhys.85.471}.

A critical road-block for unleashing the potential that optical trapping and
levitation can achieve is understanding how the spatial profile of the optical
trap affects the quality of the trap and controlling it. For example, in
traditional optical tweezers, where microparticles are trapped in aqueous
solution and motion is overdamped, optomechanical stiffness has been enhanced by
spatially modulating the trapping field using iterative algorithms
\cite{taylor2015enhanced,butaite2024photon,doi:10.1021/acsphotonics.3c01499}. In
optical levitation, there exists a theoretical framework for cooling objects of
arbitrary shapes \cite{PhysRevLett.130.083203} that could allow to achieve
unexplored quantum regimes of macroscopic systems. However, in order to
implement them, there is an experimental need to control with precision the
wavefront of the trapping beam to exploit the full numerical aperture (NA) of
the system. Unfortunately, we still have a limited understanding of the effect
of the unavoidable experimental imperfections and their corresponding effect on
the forces in the transversal and longitudinal directions with respect to the
optical trapping beam propagation. One of the main concerns in this regard are
optical aberrations caused by optical elements or misalignment, which affect the
shape of the wavefront and subsequently reduce the performance of these systems
\cite{Yu2025CorrectionOW, Schkolnik:2014kza}. Often the trapping potential gets
distorted by optical aberrations and leads to the reduction of usable numerical
aperture (NA), which can considerably decrease the trapping efficiency. In
addition to that, aberrations introduce nonlinearities in the system, and the
motion of a damped oscillator becomes more complex in comparison to a simple
harmonic motion \cite{PhysRevLett.128.213601}. In order to compensate these
aberrations, wavefront shaping devices, such as Spatial light modulators (SLM)
or Digital Micromirror Devices (DMD), can be used \cite{vcivzmar2011optical}.
The proper wavefront shaping is critical for two reasons: 1) The effect of
optical aberrations on arbitrary particles is not well understood theoretically,
and 2) different figures of merit can be used to iteratively improve an optical
trap. In this last respect, optical levitation gives us the advantage of
observing the real-time dynamical properties (i.e. CoM motion) of the levitated
particles. The three-dimensional resonance frequencies of the CoM motion of the
particle appear as the pressure level is reduced and the particle reaches an
underdamped regime. Therefore, in this case, it is usual to try to compensate
for the optical aberrations by using the trapping frequencies as observables.

\begin{figure}
\centering
\includegraphics[width=3.4in]{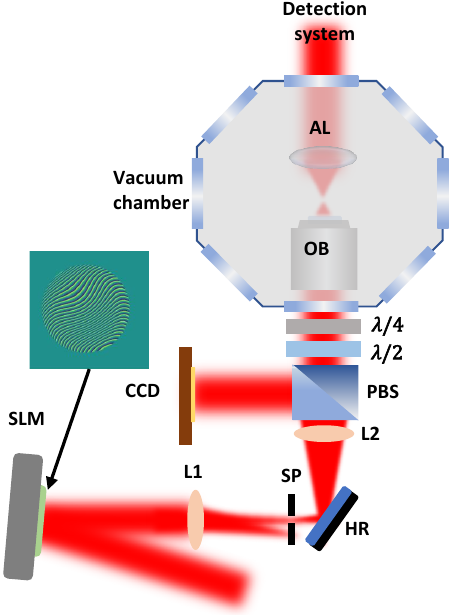}
\caption{\textbf{The schematic diagram of our experimental setup}: Starting from
  the bottom-right: A 1064 nm laser is collimated and expanded to the
  approximate diameter of the SLM. The holograms programmed on the are used to
  diffract the beam and correct its wavefront. Then a 4f system filters and
  collimate the first diffracted order into the chamber microscope objective OBJ
  inside the vacuum chamber. An aspheric lens collects the scattered light and
  send it to the detection system.}
\label{fig:setup}
\end{figure}

In this work, we demonstrate a method to solve the two main problems stated
above for optical trapping optimization of spherically shaped particles. To
achieve this, we have experimentally implemented a levitation set-up for single
silica nanoparticles by tightly focusing an infrared laser beam with a high NA
objective lens in vacuum. The trapping scheme of our system is in the direction
of gravity, as shown in Fig.\ref{fig:setup}. Here, gravity compensates for some
of the photophoretic and scattering forces, so that the gradient
force-responsible for pulling the particle towards the focus, becomes dominant
\cite{Ashkin:86}. We use a phase-only liquid crystal on silicon (LCoS) SLM in an
off-axis configuration to shape the wavefront of our trapping field. To optimize
the hologram we trap a particle and reduce the pressure to a few hPa, then we
record the Power Spectral Densities (PSDs) of the particle and measure the
resonant frequencies ($\Omega_x, \Omega_y, \Omega_z$) experimentally. On the one
hand, with this set-up we fully explore the role of aberrations on the optical
trap, by independently controlling different wavefront changes using Zernike
polynomials. These functions are commonly used to represent phase aberrations in
optical systems, as they form an orthonormal base over the circular pupil on
which they are defined \cite{Antonello2015,Noll1976}. On the other hand, in
order to theoretically understand the optical parameters, we use Generalized
Lorentz-Mie Theory (GLMT) to theoretically calculate the optical forces,
frequencies, and ratios for a range of particle sizes with fixed beam power,
beam diameter, and numerical aperture of the objective lens. This is a
theoretical framework that has been shown to give a complete model for complex
optical traps in aqueous media \cite{doi:10.1021/acsphotonics.3c01499,
  Gomez2025}. Upon comparing the experimental and theoretical results, we
provide a full description of optical aberrations and a method to control or
compensate for them for different experimental purposes.

We exploit Zernike polynomials in our system by programming the SLM and by
applying different holograms with different values of the coefficients of
Zernike polynomials. In this way, we can systematically explore the role of
structured wavefronts in optical levitation. In particular, we observe that
wavefront shaping has an enormous impact on the CoM motion frequencies and,
importantly, on their ratios. In this manner, we can assess what the optimal
configuration is depending on the purpose of the experiment. For example, we
provide a simple recipe to calculate the ratios theoretically and implement it
experimentally in order to obtain the best possible optical trapping beam.
Importantly, we have noticed that this condition does not provide the maximum
transverse trapping stiffness. This effect is consistent with recent results by
Kleine \emph{et al} \cite{kleine2025wavefrontshapingscatteringforces}who
presented a method to increase the trap stiffness by shaping the wavefront of
the trapping beam, also using Zernike polynomials. In their case, they use a
weighted sum of the three frequencies of the CoM as an objective function to
maximize. Here, instead, we propose that the ratio of the transversal to
longitudinal frequencies is a better figure of merit to assess the quality of
the trapping beam, as this figure of merit can be used independently of the
laser power and particle size. Furthermore, we show that, in our experiments,
the optimal beam shape is achieved by maximizing the longitudinal frequency. Our
results provide a comprehensive overview of the role of aberrations in optical
levitation and also a recipe for achieving optimal beam configurations,
depending on the experimentalist needs.

\section{Theory}
The forces in an optically levitated particle are divided into two main parts,
the scattering force (non-conservative) and the gradient force (conservative)
\cite{jin2024towards}. The scattering force on a particle has the same direction
as the laser's propagation, while the gradient force provides a restitutive
force directed towards the equilibrium position. The resulting dynamics
corresponds to that of a Brownian motion in a harmonic potential. The degrees of
freedom of CoM motion have the frequencies $\Omega^{(i)} =
\sqrt{k_\mathrm{trap}^{(i)}/m}$, where $m$ is the mass of the particle and
$k_\mathrm{trap}^{(i)}\ $, for $i=x, y, z$, are the elastic constant of the
potential well, or trapping stiffness. Given that the elastic constants are
proportional to the trapping beam intensity, maximizing the trapping frequencies
can be done by either optimizing the beam shape or by simply increasing the
optical power. On the other hand, as their ratio is independent of the optical
power, we will see that the ratio of these frequencies allows us to infer the
tightness of the focused beam.

Let us start considering a particle in the Rayleigh regime (or dipolar regime),
that is, when the particle diameter is much smaller than the wavelength of the
trapping beam. The gradient force in that situation is given by
\cite{gieseler2014dynamics}
\begin{equation}
\vec{F}_\mathrm{grad}(\vec{r}) = \frac{\alpha}{4} \nabla I_0(\vec{r}),
\label{eq:force-gradient-dipolar}
\end{equation}
where $\alpha$ is the effective polarizability of the dielectric particle, and
$I_0(\vec{r})$ is the trapping beam intensity. In order to get some insight on
the behavior of the frequency ratios, let us review the simpler case of
$I_0(\vec{r})$ corresponding to a Gaussian beam in the paraxial approximation.
In that case, it can be shown that the trap stiffnesses in the transverse and
longitudinal directions are \cite{Gieseler2012, Harada1996}:
\begin{eqnarray}
k_\mathrm{trap}^{(i)} & =& \alpha E_0^2/w_0^2  \qquad \quad \mathrm{for\ \ } i=x, y \nonumber \\
k_\mathrm{trap}^{(z)} &= &\alpha E_0^2/(2 z_R^2),
\label{eq:k-dipolar-paraxial}
\end{eqnarray}
where $E_0$ is the electrical field amplitude in the focus, $w_0$ is the beam
waist and $z_R= \pi w_0^2/\lambda$ the Rayleigh distance, with $\lambda$ being
the optical wavelength in free space. Thus, the ratio of the resulting
frequencies allows for a direct estimation of the ratio of beam size in the
longitudinal and transverse directions. In addition, within this approximation
the estimation is independent of both the particle size and the power of the
trapping beam. In particular, the frequency ratios are
\begin{eqnarray}
\frac{\Omega_i}{\Omega_z} = \frac{\sqrt{2}}{\mathrm{NA}} \frac{1}{\filling}\qquad \mathrm{for\ } i=x,y
\label{eq:freq-dipolar-paraxial}
\end{eqnarray}
where $NA$ is the numerical aperture of the focusing lens, while $\filling$ is
the Filling Factor, defined as the ratio between the $1/e^2$ input beam diameter
and the back aperture of the objective lens.

\begin{figure}
\centering
\includegraphics[width=3.4in]{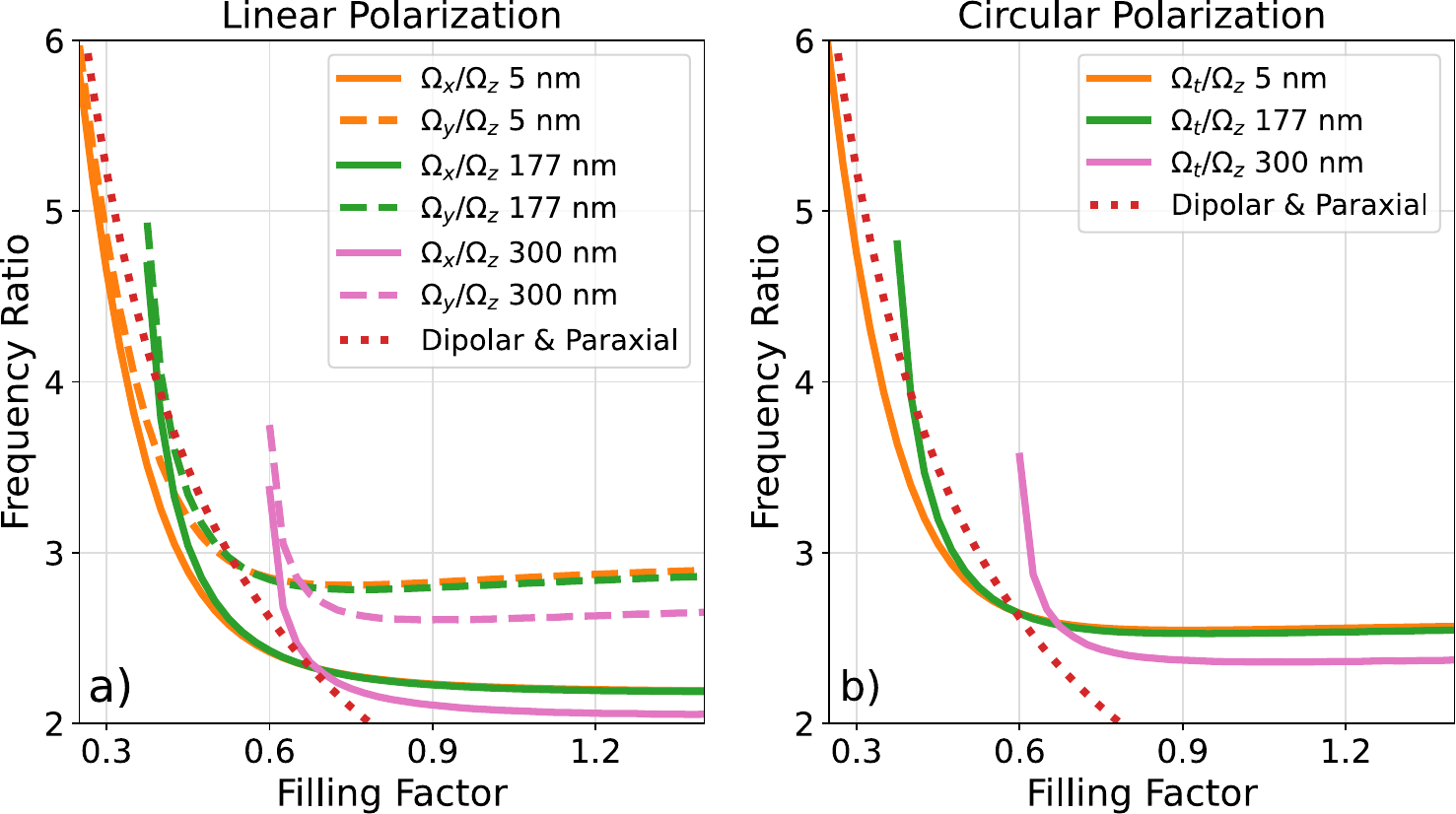}
\caption{Theoretical simulation of of back aperture filling of the objective
  lens ($NA=0.9$) in linear (a) and circular (b) polarization with respect to
  the ratios of the frequencies of the multiple particles by GLMT.}
\label{fig:simulated-ratio-vs-filling}
\end{figure}

Although equations \eqref{eq:k-dipolar-paraxial} and
\eqref{eq:freq-dipolar-paraxial} are essential to understand how to use the
information of the CoM of the trapped particle to estimate the quality of the
focus, they start to lose accuracy in the case of a tightly focused beam. When a
linearly polarized laser beam is tightly focused, the symmetry of the beam
breaks due to the vector nature of light \cite{novotny2012principles}, resulting
in different transverse frequencies in the direction of the polarization and
orthogonal to the polarization. Furthermore, these expressions for the ratios
are only valid for small particles.

In order to avoid these limitations, let us analyze the frequency ratios using
GLMT, a theory that takes into account higher multipolar excitations of the
trapped particle and is suitable for describing optical forces in all size
regimes of the particle, from the Rayleigh approximation up to ray optics
approximation \cite{10.1063/1.343813}. To calculate the optical forces and
trapping frequencies of the particle, we use the Multipolar Optical Forces
Toolbox (MOFT) \cite{Gomez2025}\footnote{Code available in the Github repository
\href{https://github.com/QNanoLab/MOFT}{MOFT}}. We performed simulations for
silica particles of sizes ranging from $5\ \mathrm{nm}$ to $300\ \mathrm{nm}$
trapped with a Gaussian beam focused by a lens of $NA=0.9$ in a vertical
configuration. Fig.\ref{fig:simulated-ratio-vs-filling} shows the predicted
frequency ratios as a function of the Filling Factor for different particle
sizes ranging from $5\ \mathrm{nm}$ to $300\ \mathrm{nm}$ of diameter. On the
one hand, for a linearly polarized input beam
(Fig.\ref{fig:simulated-ratio-vs-filling}a) due to the difference of the focused
beam waist in the directions along the polarization ($x$) and orthonormal to it
($y$), there are two relevant frequency ratios $\Omega_x/\Omega_z$ (solid line)
and $\Omega_y/\Omega_z$ (dashed line). When using circular polarization
(Fig.\ref{fig:simulated-ratio-vs-filling}b), the preserved cylindrical symmetry
gives rise to a single transverse frequency with a ratio $\Omega_t/\Omega_z$
that is between the $\Omega_x/\Omega_z$ and $\Omega_y/\Omega_z$ of the linear
polarization case.

For low filling factors, the expected frequencies $\Omega_x$ and $\Omega_y$ for
a particle become similar, since a looser focus makes the waists $w_x$ and $w_y$
to match. However, it is readily apparent that the dipolar and paraxial
approximation of equation \eqref{eq:freq-dipolar-paraxial} (dotted line in
Fig.\ref{fig:simulated-ratio-vs-filling}) is a good approximation \emph{only}
for the lower filling region of the $5\ \mathrm{nm}$ particle. This can be
understood considering the overlap between the multipolar content of the Mie
coefficients of the spherical particle and that of the trapping beam: a loosely
focused beam features a higher number of multipolar orders, which in turn couple
to the higher multipolar orders associated with a bigger particle
\cite{Zambrana-Puyalto2012, Molezuelas-Ferreras2023,
  doi:10.1021/acsphotonics.3c01499}. It should be noted that for filling factors
larger than $0.5$, the difference in frequency ratios for particles with
diameter smaller than $177\ \mathrm{nm}$ are negligible in a 1064nm trapping
laser. For particles bigger than that, as is the case for the ratio curves for
the 300 nm particle, the curves differ from the expected paraxial dipolar regime
in all the filling factor range.

Thus, these predicted results allow us to assess the quality of the focused beam
by measuring the CoM frequency ratios, a quantity that is independent of the
power of the laser. Furthermore, for small particles (smaller than 177 nm in
diameter), the expected ratio is independent of the actual particle size. In the
case of larger particles, the expected ratio can be found using the MOFT code.

\section{Experimental setup}
The schematic diagram of our experimental setup is shown in Fig.\ref{fig:setup}.
A 1064 nm laser is expanded and collimated with a telescope to fill most of the
SLM screen. A combination of a blazed phase diffraction grating hologram and a
phase mask is applied to the SLM, and the first diffracted order containing the
structured wavefront is filtered in the middle of the $4f$ system by a spatial
filter and collimated by the second lens of the $4f$ system. The collimated beam
passes through a PBS (polarizing beam splitter), one part of the beam is used to
assess the size of the beam under different wavefront patterns by using a CCD
camera at the same distance as the back-aperture of the objective lens and the
other part passes through a set of waveplates (\textbf{WPs}) to control the
polarization of the trapping laser before entering the vacuum chamber. Another
aspheric lens is used to collect and collimate the scattered and directly
transmitted light from the trapped particle, which is then directed to the
detection system. More experimental details are given in (\textbf{Supplementary Document}).

\section{Methodology}
In our experiment, we used commercially available 177 nm diameter amorphous
silica nanoparticles (Bangs Labs), that were loaded into the trap using an
Omicron nebulizer at ambient pressure. Then the pressure is reduced to a few hPa
where the gas damping becomes negligible. We record the power spectral density
(PSD) of the particle in both a linearly and a circularly polarized beam. As
explained above, our system allow for the collection of the PSDs while
structuring the wavefront of the trapping beam. From the PSDs we extract the
information of the resonant frequencies and compute the ratio between the
transverse and longitudinal frequencies.

In a typical experiment, we start with an uncorrected beam where the ratios do
not match the theoretical results. Then, we will explore different wavefront
corrections. This is done systematically with the use of superpositions of
Zernike polynomials, whose coefficients are sweeped simultaneously until we find
a configuration which match the theoretically predicted frequency ratio values.
We are using the same definition for the Zernike polynomials as in Ref.
\cite{Antonello2015}, with a normalization factor that corresponds to minimum
and maximum values of the phase of $\pm \pi$. The explicit expressions of the
polynomials used are shown in (\textbf{Supplementary Document}). The pupil position and
radius of the Zernike polynomials mask is selected on a specific area of the SLM
and match the size of the beam impinging the SLM (independently measured by a
raster scan on the SLM). Importantly, we keep track of the
variations of the filling factor due to the propagation of the structured
wavefront beam, using an imaging system.

\section{Results and discussion}

\begin{figure}
\centering
\includegraphics[width=3.4in]{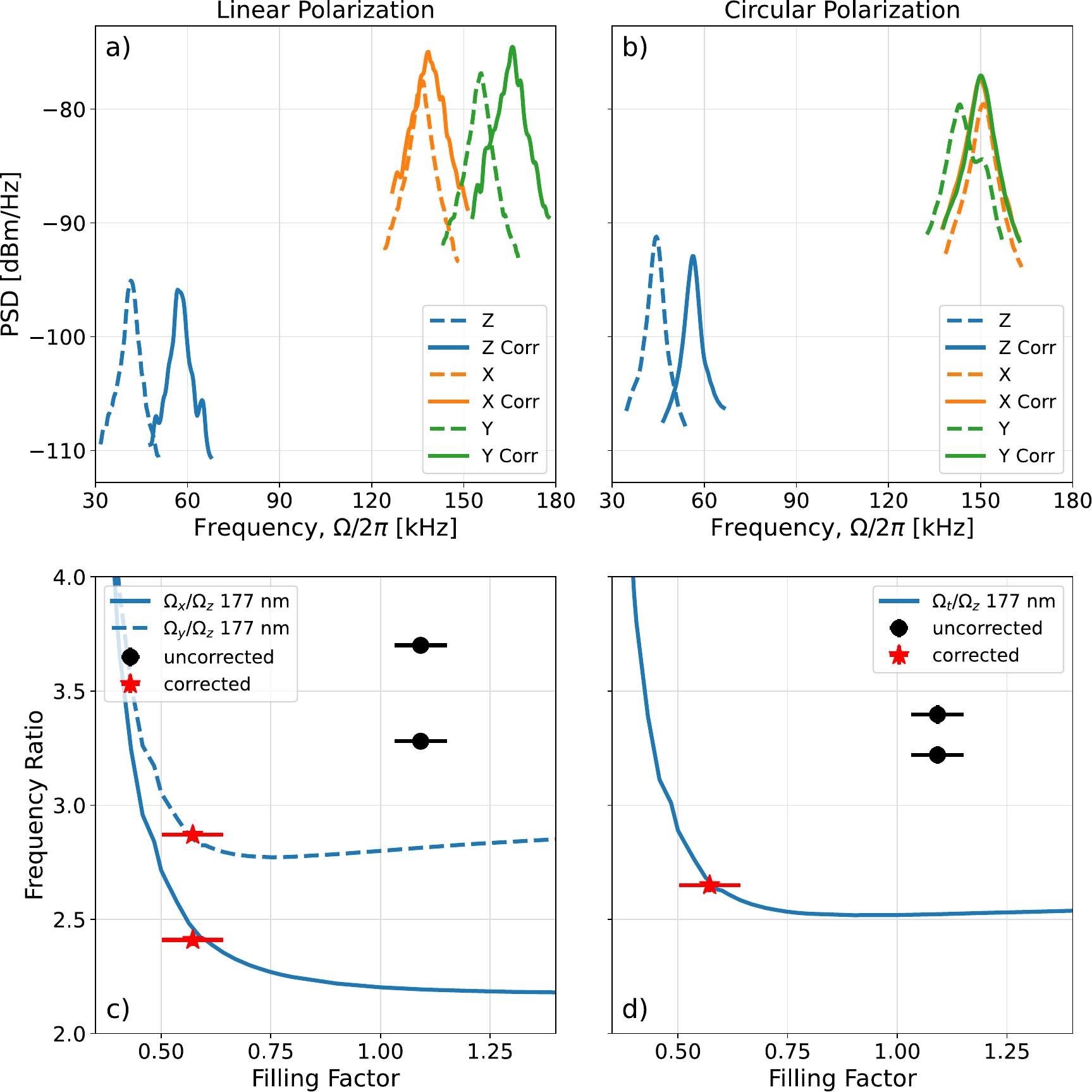}
\caption{(a)-(b) represents the \textbf{ Power spectrum density} of levitated
  nanoparticle. PSD in (a) linear and (b) in circular polarization, the dashed
  lines represent uncorrected and the solid lines are in corrected beam. (c)-(d)
  is the \textbf{ Frequency ratios} of levitated nanoparticle. The solid lines
  represents the theoretical prediction of back aperture filling of the
  objective lens in linear (c) and circular (d) polarization with respect to the
  ratios of the frequencies of the particle. The black circular dots in (c) and
  (d) represents the CoM motion ratios in uncorrected beam profile. The red star
  in (c) and (d) represent the CoM motion ratios in the corrected beam profile.}
\label{fig:spectra-comparison}
\end{figure}

\begin{figure}
\centering
\includegraphics[width=3.4in]{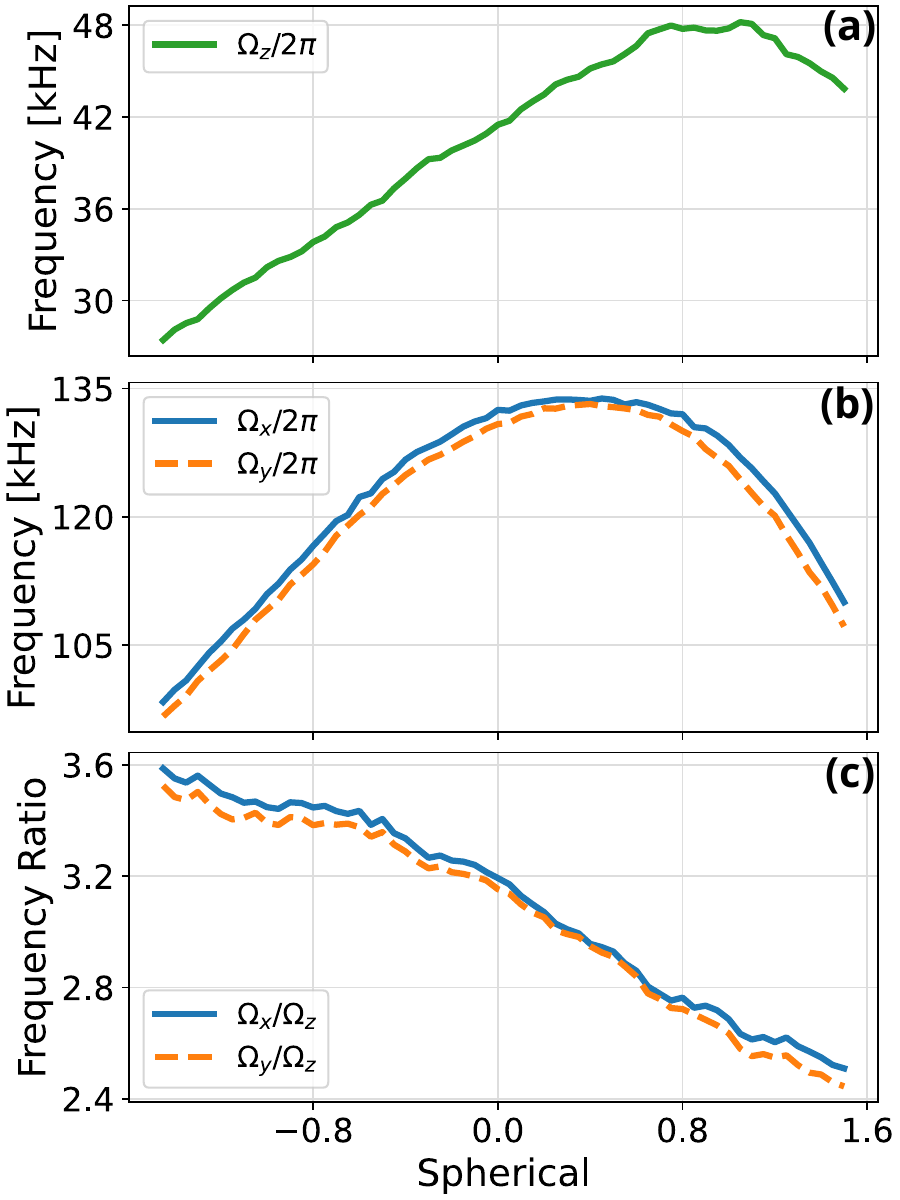}
\caption{\textbf{Sweep of Spherical in circular polarization}. The green line in
  (a) represents the variation of frequency along the longitudinal axis
  ${\Omega_z}$, (b) shows the change in frequencies along transversal axis
  ${\Omega_x}$, ${\Omega_y}$. And (c) represents the ratios between transversal
  and longitudinal directions ${\Omega_x}/{\Omega_z}$ and
  ${\Omega_y}/{\Omega_z}$, respectively.}
\label{fig:spherical-sweep}
\end{figure}

Let us start by displaying a typical resulting PSD with the resonance
frequencies when trapping a particle with a focused Gaussian beam without any
wavefront correction. These can be observed in the dotted line curves of
Figs.~\ref{fig:spectra-comparison} (a) for a linearly polarized beam and (b) for
circular polarization. In both cases, three different peaks can be clearly seen,
the lowest frequency (between 30 and 40 kHz), corresponding to the longitudinal
oscillation of the particle ($z$-axis), and the other two around 120-160 kHz
corresponding to the transversal frequencies ($x$-axis and $y$-axis). In
circular polarization it can be observed that the two transversal peaks are
closer than in the linear polarization case. Although these results are
qualitatively consistent with the theoretical description above, there are some
quantitative elements that pinpoint the distortive effect of the aberrations on
the beam, possibly caused by different optical elements and, in particular, from
misalignments and imperfection of the focusing microscope objective. The most
noticeable one is that in the circular polarization case, the transversal
frequencies are displaced from each other, meaning that we have lost the
cylindrical symmetry of the beam. A more subtle indication is that the ratio of
the transversal to longitudinal frequencies is far away from the expected ratio
for this particle size, numerical aperture and filling factor of the system.
This can be observed in Fig.\ref{fig:spectra-comparison} (c) and (d), where we
plot the two frequency ratios for both polarization cases with a black dot and,
for comparison, we plot the theoretically expected ratios calculated above
(continuous and dashed lines).

We set ourselves to the task of optimizing the trapping beam. To that objective.
we progressively changed the correcting factors of the Zernike polynomials
\cite{Antonello2015, Lakshminarayanan2011} using the SLM. We start the
optimization procedure by first correcting for the astigmatism which is the
lowest order asymmetric Zernike polynomial. This allows us to compensate for the
differences of the transversal frequencies in the circular polarization case.
Once the cylindrical symmetry of the system is restored and the $x$ and $y$
frequencies are at the same position, we can proceed scanning with the other
Zernike polynomials. Even though we had access to up to 27 different Zernike
polynomials, we realized that, typically, only the lowest order ones have an
appreciable impact on the trapping frequencies, mainly the so-called
\textit{Defocus}, \textit{Spherical} and to a lesser extent \emph{Coma and
Astigmatism}. The \textit{Defocus} has a quadratic behaviour with radius and
corrects the overall focus shift and the \textit{Spherical} Zernike polynomial
corrects for focus discrepancies in central and peripheral rays due to its
quartic nature. The resonance frequencies produced by the optimized beam profile
are shown by solid lines in Figs.~\ref{fig:spectra-comparison} (a)-(b). Now, it
is observed that the frequencies agree also quantitatively with the expected
results: the transversal frequencies are at the same position with circular
polarization and the $z$-axis frequency is the same for linear and circular
polarization. The ratios of the frequencies shown in
Fig.\ref{fig:spectra-comparison}(c)-(d), fall exactly onto the predicted curve,
once we take into account the difference in the filling factor, which we
independently measure. This ensures that our correction process has achieved an
optimal beam. However, it is interesting to notice that although the $z$-axis
frequency has increased around 15-20$\%$, the optimization process that we
follow does not have such a great impact on the transverse frequencies.

In order to further investigate and understand the role of aberrations on the
quality of the beam and the optimization of the trapping frequencies, we have
performed an exhaustive study of the most important aberrations of the system.
Starting with the \textit{Spherical} aberration, in Fig.
\ref{fig:spherical-sweep} we present typical experimental results for circular
polarization. The effect of sweeping  the \textit{Spherical}
aberration on the \textit{z-axis} frequencies is presented in (a). One can
observe that we reach a maximum frequency at a particular value of the
polynomial coefficient (around 0.9 in this case). The shape of the curve
presents a linear increase of the frequency, which tops at a value of around
50$\%$ of the lowest value. In Fig.~\ref{fig:spherical-sweep} (b) we present the
corresponding behavior for the transverse frequencies. When comparing this curve
with the one above, we see that this curve has a more pronounced quadratic
behavior, with a smoother peak, which reaches around 30$\%$ of the minimum
frequency value. Also, it is quite observable that the \textit{Spherical}
aberration coefficients that maximize the longitudinal and transverse
frequencies are different. In an automatic optimization algorithm, this may
produce an ambiguity in reaching an optimal point that can be solved by using as
a figure of merit a particular weighted value of the frequencies for different
axis \cite{kleine2025wavefrontshapingscatteringforces}. However, from the
behavior of the frequency ratios (in Fig.~\ref{fig:spherical-sweep} (c) ), it is
observed that the point that corresponds best to the theoretically expected
value is closer to the peak of $z$-axis frequency.

\begin{figure*}
\includegraphics[width=5in]{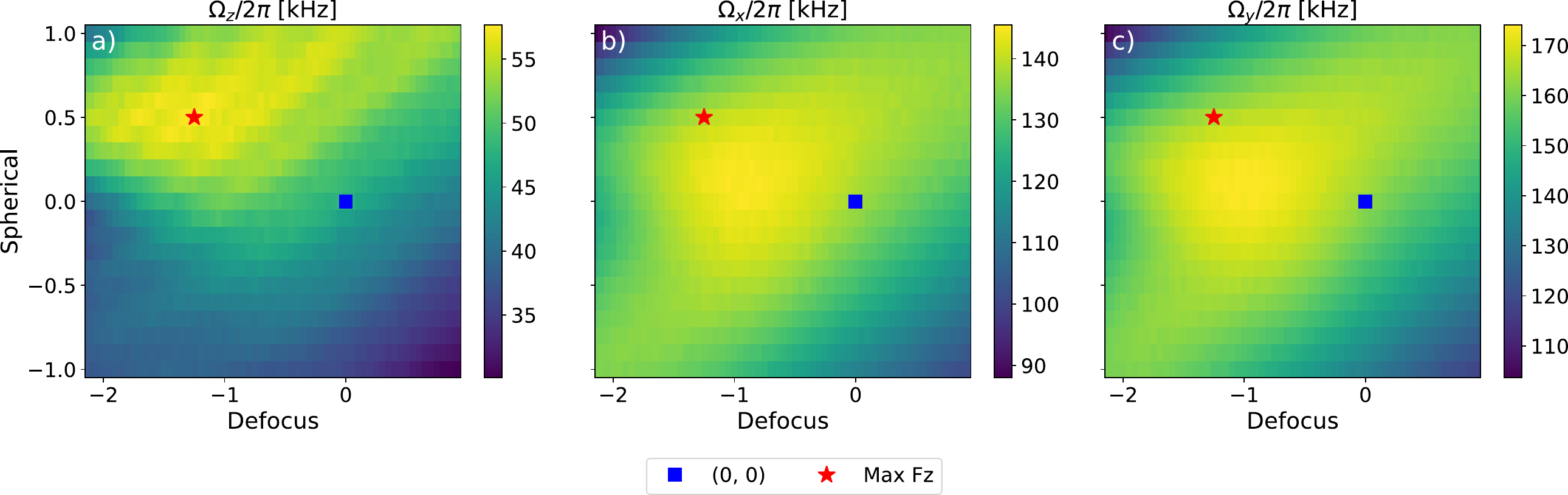}
\caption{Maps of the CoM frequencies for the (a) z, (b) x and (c) for for a grid
  sweep of the \emph{Spherical} and \emph{Defocus} parameters in linear
  polarization. The fixed values of \emph{Astigmatism X} and \emph{Astigmatism
  D} were selected to restore the cylindrical symmetry in circular polarization,
  while the fixed values of \emph{Coma X} and \emph{Coma Y} where selected to
  maximize $\Omega_z$, resulting in an small increase of around
  $3\ \mathrm{kHz}$.}
\label{fig:spherical-defocus-map}
\end{figure*}

In order to confirm that these results are not due to having reached a local
maxima in the full space of Zernike polynomials, we have produced a two
dimensional map of the lowest order coefficients. The frequencies obtained are
shown in Fig.\ref{fig:spherical-defocus-map}. We have used linear polarization
in this case in order to show that our results hold for any polarization, and
also how different the behaviour of the $x$ and $y$-axis frequencies is in this
case. The first feature that can be observed in these figures is that the
\textit{Defocus} and \textit{Spherical} aberrations show a coupled behaviour,
i.e. the frequency patterns show a diagonal pattern, indicating that in an
optimization algorithm these aberrations must be treated together and cannot be
optimized in sequence. The $z$-axis frequency in particular shows some shallow
local minima and a very smooth change when moving along one of the ridges that
are apparent when we change both aberrations at the same time. Also, no
appreciable differences can be observed when comparing the $x$ and $y$-axis
behaviour, except for an overall higher value in the polarization direction.
Altogether, it is clear on the map that the two transversal frequencies have a
maximum at a position different from the $z$-axis frequency. The optimal point
(the one coinciding with the optimal ratio in
Fig.\ref{fig:spectra-comparison}(c)-(d) is also represented by a star in the map
and clearly coincides with the global maximum of the $z$-axis frequency. These
results are valid when we also correct the higher order aberrations, as their
effect is much smaller, but they allow us to improve the performance of the
system by around 5$\%$.

This systematic study of the impact of optical aberrations in levitation
experiments indicates that an experimentalist can choose between maximizing the
transverse frequencies at the cost of a shallower trap in the longitudinal
direction. Also, they show that the experimental coefficients that are most
compatible with a wavefront corrected beam, according to a full theoretical
model (for non-dipolar particles and vectorial beams), correspond to optimizing
the longitudinal frequency (at the cost of not maximizing the transversal
frequencies).

\section{Conclusion}
In conclusion, we have presented a method to assess the quality of the trapping
beam which uses frequency ratios of the particle's center of mass motion that is
independent of the trapping optical power and, for sufficiently small particles,
also independent of the particle's diameter. Furthermore, we presented a study
of how different structured wavefronts can optimize the optical trapping in
different manners. Our results indicate that the wavefront correction that
provides the best possible beam (as indicated by the transversal to the
longitudinal frequency ratios) would also achieve the maximal longitudinal
frequency. On the other hand, in order to maximize the transversal frequency,
one has to use a slightly aberrated beam.

In order to achieve these results, we theoretically calculated the trap
stiffness and the ratios of their corresponding frequencies by GLMT as a
function of the microscope objective filling factor, which allows us to predict
the ratio between the transverse and longitudinal frequencies for different
tightnesses of the focusing. For particles smaller than $177\ \mathrm{nm}$ these
ratios are independent of the actual particle size. For bigger particles,
although the expected ratios change, they can be reliably estimated evaluated
using GLMT.

Thus, we present a method that provides an easy way to evaluate the beam
quality, and a simple recipe to calculate the ratios theoretically and implement
it experimentally to correct the potential aberrations of the beam, starting
from the astigmatism and continuing with the lowest order symmetric Zernike
polynomials. These corrections are of great importance and paves the way for a
well-optimized trapping potential for complex materials
\cite{PhysRevLett.129.023602}. This system can be used to mold the beam profile
according to our own needs for many quantum applications, i.e. stable levitation
of different materials in high vacuum \cite{Kuhn2017a}, materials embedded with
emitters \cite{jin_quantum_2024}, quantum superposition experiments and decrease
of the backaction \cite{2yzc-fsm3}.

\section*{Funding}
Departamento de Educacion, Basque Government. IKUR Quantum Technologies. PTI-01 Quantum Technologies, CSIC. Proyecto PID2022-143268NB-I00 financiado por MCIN.

\section*{Acknowledgment} Shah Jee Rahman, Quimey Pears Stefano and Gabriel
Molina Terriza acknowledge support from the CSIC Research Platform on Quantum
Technologies, from IKUR Strategy under the collaboration agreement between
Ikerbasque Foundation and DIPC/MPC on behalf of the Department of Education of
the Basque Government. 

\section*{Disclosures} The authors have no conflicts to disclose.

\section*{Data Availability}

Supporting data is available upon reasonable request from the authors.

\section*{Supplementary information}
See section Supplementary Document at the end of this document.


\bibliography{References}

\onecolumngrid
\section{Supplementary Document}

\subsection{Details of the experimental setup}
\begin{figure}[ht!]
\centering \includegraphics[width=3.4in]{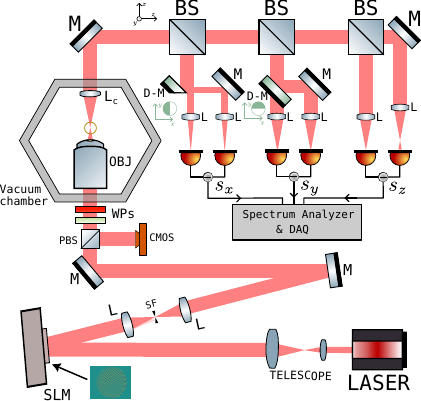}
\caption{\textbf{The schematic diagram of our experimental setup}: Starting from
  the bottom-right: A tuneable laser (1040-1075nm) followed by a half waveplate
  and PBS to control the intensity. Then an AOM for shifting the frequency and a
  beam expander. The SLM is used to diffract the beam and shape the beam
  profile. Then a 4f system to filter and collimate the first order beam all the
  way into the chamber. An aspheric lens to collect the scattered light and send
  it to the detection system.}
\label{fig:setup-detail}
\end{figure}

Fig.\ref{fig:setup-detail} illustrates the experimental setup of our optically
levitated spheres in a structured beam. We are using a fiber-coupled Toptica DL
TA Pro tuneable laser (1040 - 1075 nm) with Gaussian mode TEM00 at a fixed
wavelength of 1064 nm. After the fiber collimation lens, the beam propagates
through a half-wave plate $(\lambda/2)$ and then subsequently passes through a
polarizing beam splitter (PBS) cube to control the intensity of the laser (power
of the beam). The linearly polarized beam after the PBS passes through an
acousto-optic modulator (AOM) model (MT80-A1.5-1064) and the frequency is
shifted by a mount of 80 MHz for future experiments. The shifted beam is then
expanded (7mm) and collimated to fill most of the area of the spatial light
modulator (EXULUS-HD3, 650-1100nm). The first order of the diffracted beam from
the SLM passes through the first lens of the 4\textit{f} system; we spatially
filter the beam at the focus point and is subsequently collimated by the second
lens. The collimated beam then passes through another PBS for imaging the beam
to measure the diameter of the beam at a distance equal to that of the
back-aperture of the objective lens by a CCD (CS165MU), and the other half
passes through a combination of a half-wave $(\lambda/2)$ and a quarter-wave
$(\lambda/4)$ plate to control the polarization of the laser of the trapping
beam. The trapping beam is then vertically directed into the vacuum chamber
(MCF600-SphOct-F2C8) and tightly focused by a microscope objective lens with a
high numerical aperture (CFI Plan NCG 100X, WD = 1 mm, NA = 0.9) in free space.
The total power of the trapping laser before entering the chamber is 172 mW and
we have losses of about 58\% in the window and objective lens. The power in the
trapping region is about 72 mW. The silica nanoparticles (SiO$_2$) of around 177
nm are in aqueous solution, which are first diluted in very pure ethanol at a
concentration of $1.862 \times 10^{10} \text/mL$ (making a ratio of 1:1000) and
then sonicated for approximately 15 minutes.

We used another high NA aspheric lens (C330TMD-C, WD = 1.8 mm, NA=0.7, ARC
1050-1700 nm) to collect and collimate the scattered and unscattered light from
the trapped particle. The output beam is divided into three parts for the
detection of CoM motion eigenfrequencies in all three dimensions, i.e. \emph{x,
y}, and \emph{z} in a balanced photodetection scheme. We have used D-shaped
reflective mirrors to divide the beam into two equal halves for the detection of
CoM motion in the $x$ (horizontal) and $y$ (vertical) directions. Each half is
then focused by short focus lenses (\emph{f = 30mm}) on the photodiodes of the
current-subtraction detectors (PDB210C/M, $\lambda$ = 800-1700nm, bandwidth =
DC-1MHz). For the $z$ direction, a 1:2 beam splitter (BS) is used to control the
intensity of the beam. The reflected light is focused on the photosensitive area
of the photodiode. The part transmitted through the BS is focused and expanded
onto the photosensitive area of the photodetector. This makes the beam
cross-section larger than the photosensitive area of the photodiode. When the
particle gets trapped, the particle's movement in the axial direction changes
the convergence of the outgoing light. So, the intensity of the reflected light
measured by the first photodiode is related to the axial displacement of the
particle and the detector we have used for this direction is (PDB220A2/M,
$\lambda$ = 190-1100 nm, bandwidth = DC-1 MHz). We use different detectors for
the longitudinal and transverse directions and the noise floor varies in
measurement of longitudinal and transverse directions-can be observe in the main
text Fig.3 and \ref{spectra-comparison}.

\subsection{Definition of the Zernike Polynomials}
Zernike polynomials are a set of orthogonal mathematical functions defined over
a unit circle that are widely used in optics, introduced by Frits Zernike in
1934 \ref{cite-von1934beugungstheorie}, particularly in wavefront analysis and
correction, including optical trapping and levitation. Their orthogonality and
ability to represent common optical aberrations make them ideal for resolving
complex wavefront distortions into understandable aberration modes. Wavefront
aberrations degrade the performance of the optical system, leading to unstable
potential traps. Correcting these aberrations requires a deformable mirror (in
our case SLM) to compensate for the aberrations through wavefront shaping.
\begin{table}[h!]
\centering
\caption{Wavefront Aberrations in Zernike Polynomials with the exact value given
  (OSA/ANSI Convention)\ref{cite-von1934beugungstheorie}}
\begin{tabular}{lllll}
\toprule
\textbf{Aberration Type} & \textbf{Zernike Term} & \textbf{Common Name} & \textbf{Wavefront Form} & \textbf{values} \\
\midrule
Defocus & $Z_2^0$ & Defocus & $\rho^2 - 1$ & -1.25\\
Spherical & $Z_4^0$ & Spherical Aberration & $\rho^4 - \rho^2$  & 0.50\\
Vertical Astigmatism & $Z_2^{-2}$ & Astigmatism-X & $\rho^2 \sin(2\theta)$ & -0.1\\
Oblique Astigmatism & $Z_2^{2}$ & Astigmatism-D & $\rho^2 \cos(2\theta)$ & 0.23\\
Horizontal Coma & $Z_3^{1}$ & Coma-X & $\rho^3 \cos(\theta)$ & -0.4\\
Vertical Coma & $Z_3^{-1}$ & Coma-Y & $\rho^3 \sin(\theta)$  & 0.4\\
\bottomrule
\end{tabular}
    \label{tab:zernike-table}
\end{table}
As Zernike polynomials are typically used to represent aberrations in optical
systems defined by circular pupils, they are defined in the disk $0 \leq \rho
\leq 1$, $0 \leq \theta \leq 2\pi$, where $(\rho, \theta)$ are the polar
coordinates. And the general equation can be factorized in the linear in

The general equation for Zernike polynomials that are defined over the disk $0
\leq \rho \leq 1$, is [\ref{cite-born2013principles}]:
\begin{equation}
Z_n^m (\rho, \theta) = R_n^m (\rho) \cdot \Theta^m (\theta),
\end{equation}
 $n$ = radial order ($n \geq 0$), $m$ = azimuthal frequency ($|m| \leq n$, $n -
|m|$ even), $R_n^m (\rho)$ is polynomial of order $n$ that encodes the radial
dependency, and $\Theta^{m}(\theta)$ encodes the angular
dependence \ref{cite-born2013principles} \ref{cite-Noll1976}. The expression that defines
$\Theta^{m}(\theta)$ is
\begin{equation}
    \Theta^{m}(\theta) = \begin{cases}
\cos(m\theta), & m \geq 0 \\
\sin(m\theta), & m < 0.
\end{cases}
\end{equation}

\begin{figure*}
\centering
\includegraphics[width=5in]{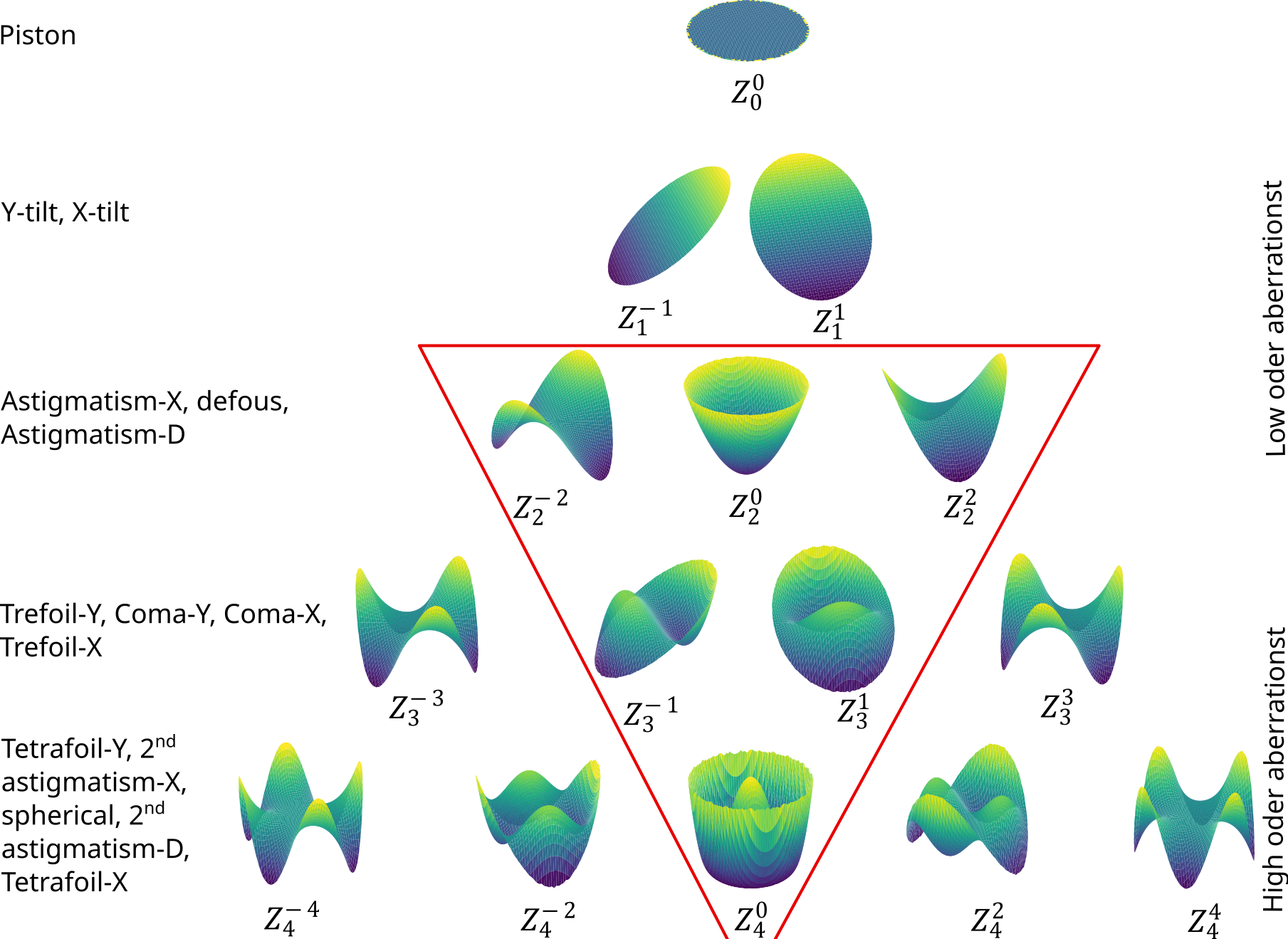}
\caption{Surface plots of Zernike polynomials with the names from left to right
  and their associated aberrations up to 4th order. The polynomials in red
  triangle are used in our experiment.}
\label{Zernike_pyramid}
\end{figure*}
Fig.\ref{Zernike_pyramid} shows typical surface plots of Zernike polynomials up
to just the 4th order, and the polynomials in the red triangle are used in this
experiment \ref{cite-Antonello:15}. The exact value and associated aberrations are
given in Table \ref{tab:zernike-table}. The most important ones in our
experiment are the cylindrically symmetric polynomials such as \emph{Defocus}
and \emph{Spherical}, and most of the aberrations are corrected for with these
two. The \emph{Astigmatisms} are used to recover the broken cylindrical symmetry
in circular polarization, and \emph{Comas} contribute as well to the overall
improvement of the system.

\subsection{Maximization of trapping frequencies}
\begin{figure*}[htbp]
\centering
\includegraphics[width=5in]{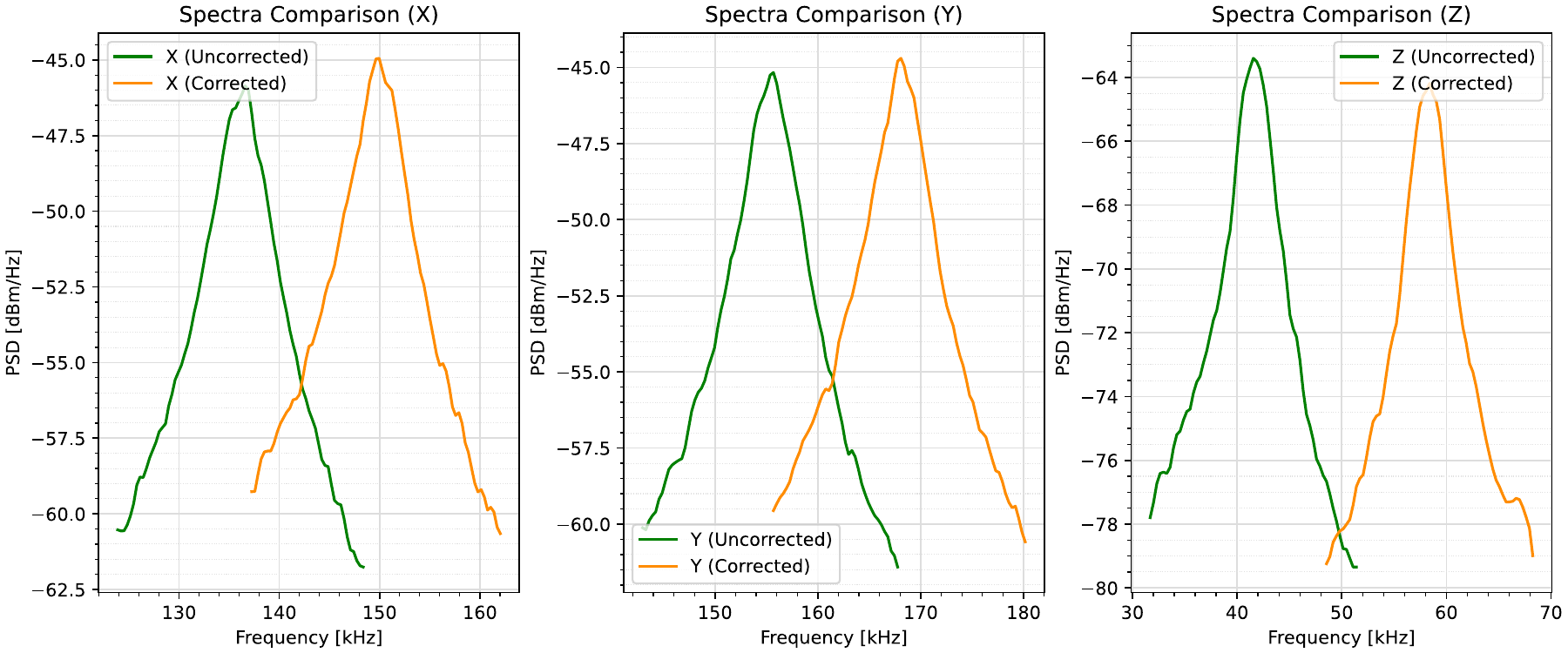}
\caption{PSDs of trapped particle in uncorrected (green) and corrected wavefront (orange) along x, y and z-axis, respectively.}
\label{spectra-comparison}
\end{figure*}
In optical levitation, higher trapping frequencies are essential for quantum
optomechanics and accurate measurement. The intrinsic confinement strength is
limited because of aberrations in the system and is the fundamental drawback of
traditional optical traps produced by a uniform laser wavefront. In this method,
we use an SLM to add structured phase patterns to the laser profile.The
symmetric phase modulation like \emph{Spherical} and \emph{Defocus} produces a
more constrictive potential well. With the same laser intensity, this technique
significantly increases the CoM motion frequencies by improving confinement in
all spatial dimensions. This technique aids in overcoming two important heating
mechanisms: backaction and photon recoil from scattered light. A more stable
mechanical system results from fewer photons interacting with the particle. With
mesoscopic objects, this development creates a clear route to the quantum
domain. Rapid decoherence caused by photon recoil is a significant element and
has been the main challenge in studies seeking to prepare quantum superposition
states \ref{cite-2yzc-fsm3}. The observation of quantum superposition states with
levitated nanoparticles becomes a more realistic objective since the photon
recoil heating that previously restricted the feasibility of such studies is
significantly suppressed.

In Fig. \ref{spectra-comparison}, we show that by playing with only symmetric
polynomials like \emph{Defocus} and \emph{Spherical}, we only increase the
stiffness of the trap on all axes of the trap. This evidently increases the CoM
motion frequency along the laser axis and in the transversal directions as well.

\subsection{Cylindrically asymmetric Zernike polynomials}
\subsubsection{Astigmatisms}
\begin{figure*}[htbp]
\centering
\includegraphics[width=5in]{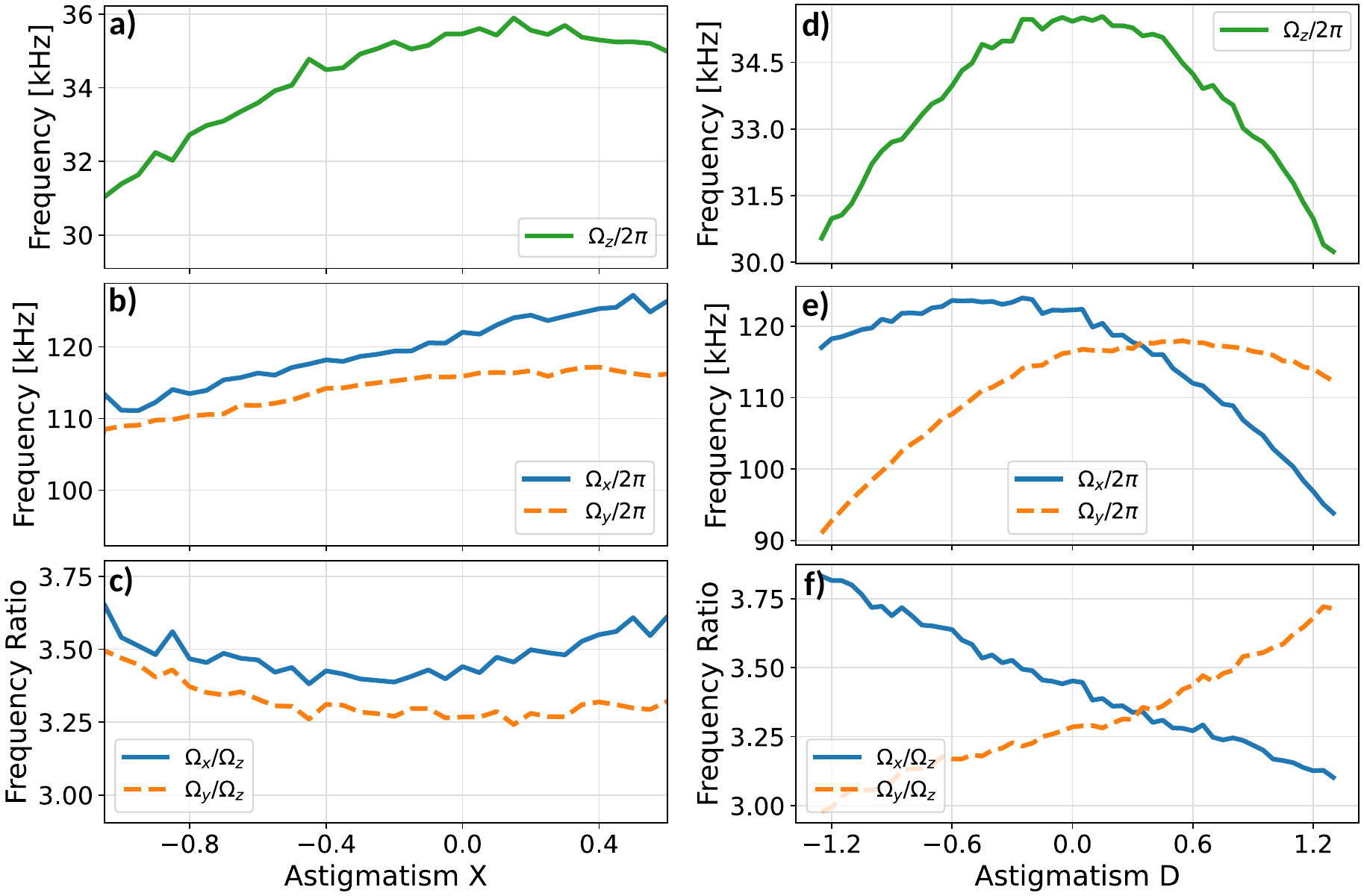}
\caption{The effects of Astigmatism-X \& Astigmatism-D on the CoM motion of
  levitated particle. The green curve in (a), (d) shows the changes in frequency
  along laser axis while sweeping the astigmatism-x and astigmatism-d,
  respectively. In (b) and (e) the blue curve represents the frequency change of
  x-axis and orange dashed represents the frequency change of y-axis in
  astigmatism-x and astigmatism-d, respectively. And lastly, (c) and (f)
  represents the ratios between (b)/(a) and (e)/(d).}
\label{fig:astigmatism-xd-sweep}
\end{figure*}

Astigmatism is an aberration in optics in which different parts of rays do not
focus at a single point. The usual cause of astigmatism in an optical system is
the lens or mirrors. Introduces asymmetry in the trapping potential, and the
stiffness varies along different axes. In a highly focused linearly polarized
beam, it is hard to point out the effects of astigmatism, as cylindrical
symmetry of the beam is already broken due to the vectorial nature of ligth.
However, in a circularly polarized beam, the circular symmetry should be
conserved and if we see that the \textit{x-axis} and \textit{y-axis} frequencies
do not match, then it becomes evident that symmetry is broken due to
astigmatism. We compensate for the astigmatisms with the SLM and recover the
circular symmetry of the beam in circularly polarized beam and then return back
to linear polarization. The main correction in our system is made with
\textit{Astigmatism-X} (Vertical Astigmatism) $Z_2^{-2}$ \textit{Astigmatism-D}
(Oblique Astigmatism) $Z_2^{2}$. In Fig. \ref{fig:astigmatism-xd-sweep}, we show
the effects of \textit{Astigmatism-X} and \textit{Astigmatism-D}. We can see
that it has far fewer effects on the longitudinal axis, and a sharp decline in
frequency can be seen while the coefficient is swept to either side. However, it
affects transverse axes much more than longitudinal ones. Increase or decrease
the separation between two transversal axes, i.e. \textit{x-axis} and
\textit{y-axis}

\subsubsection{Comas}
\begin{figure*}
\centering
\includegraphics[width=5in]{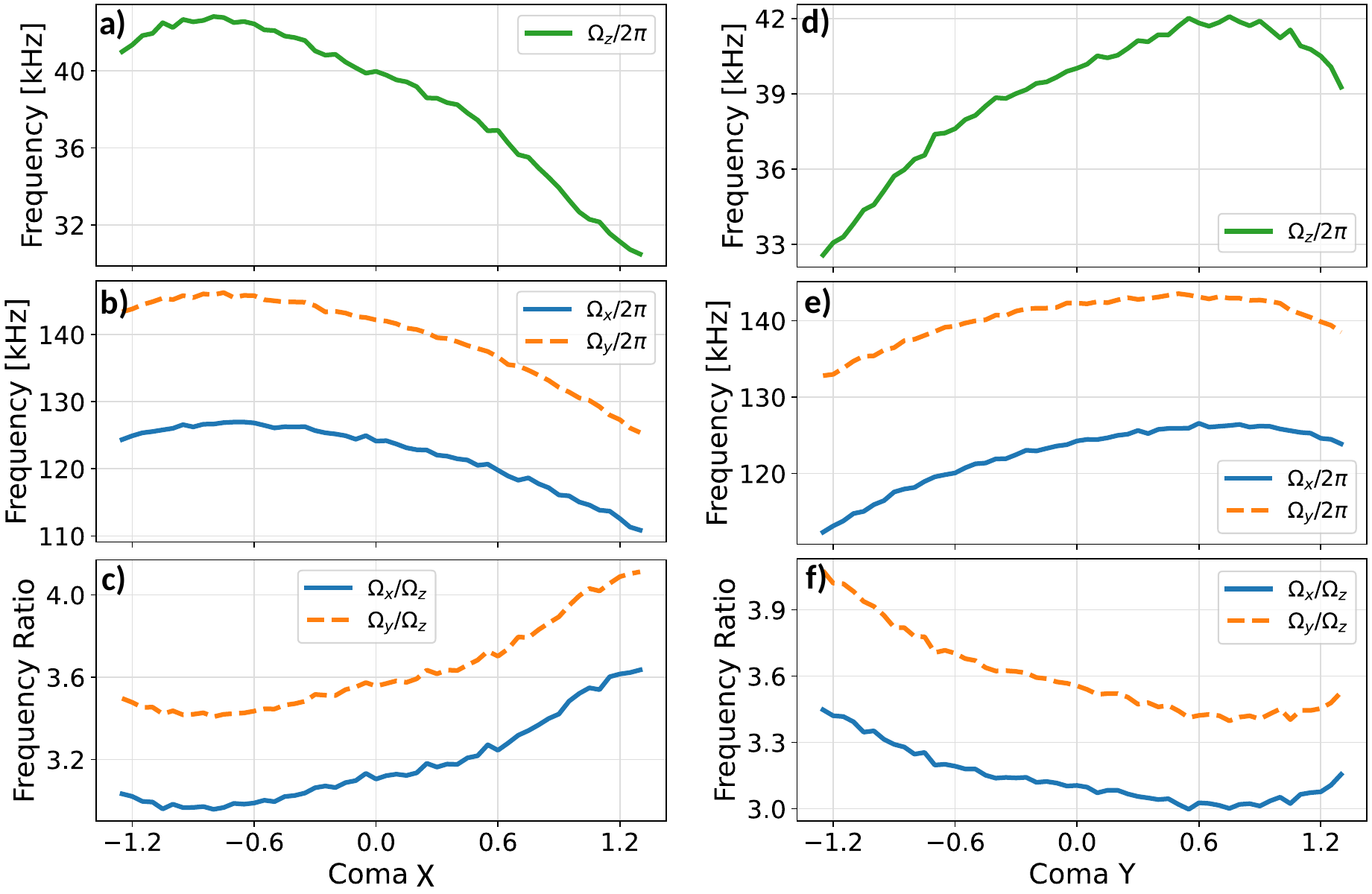}
\caption{The observed effects of Coma-X \& Coma-Y on the CoM motion of optically
  levitated particle. The green curve in (a), (d) shows the changes in frequency
  along laser axis while sweeping the coma-x and coma-y, respectively. In (b)
  and (e) the blue curve represents the frequency change of x-axis and orange
  dashed represents the frequency change of y-axis frequencies in coma-x and
  coma-y, respectively. And lastly, (c) and (f) represents the ratios between
  (b)/(a) and (e)/(d).}
\label{fig:coma-xy-sweep}
\end{figure*}

Although the effect of higher order cylindrically asymmetric Zernike polynomials
is small, Comas-X and Coma-Y have slightly noticeable effects and can increase
the longitudinal frequency up to 5\%, which suggests that Comas compensates for
aberration along the laser axis as well. In Fig. \ref{fig:coma-xy-sweep}, we
show the behavior of Comas on optically levitated nanoparticles, and it can be
seen that the improvements of frequencies in all three degrees of freedom are
very less than the \textit{Defocus} and \textit{Spherical} given in the main
text. The elimination of Comas results in a symmetric and stable potential well,
especially in vacuum, to push towards the quantum regime.

\clearpage

\subsection{References}
\newcommand{\enquote}[1]{``#1''}

\begin{enumerate}[label={[\arabic*]}]
\item Z.~von F, \enquote{Beugungstheorie des schneidenver-fahrens und seiner
  verbesserten form, der phasenkontrastmethode,}
  Physica \textbf{1}, 689--704 (1934). \label{cite-von1934beugungstheorie}
\item M.~Born and E.~Wolf, \emph{Principles of optics: electromagnetic theory of
  propagation, interference and diffraction of light} (Elsevier, 2013).\label{cite-born2013principles}
\item R.~J. Noll, \enquote{Zernike polynomials and atmospheric turbulence,}
  J Opt Soc Am \textbf{66}, 207--211 (1976).\label{cite-Noll1976}
\item J.~Antonello and M.~Verhaegen, \enquote{Modal-based phase retrieval for
  adaptive optics,} J. Opt. Soc. Am. A \textbf{32},  1160--1170 (2015). \label{cite-Antonello:15}
\item M.~Rossi, A.~Militaru, N.~Carlon~Zambon, \emph{et~al.}, \enquote{Quantum
  delocalization of a levitated nanoparticle,} Phys.   Rev. Lett. \textbf{135}, 083601 (2025).\label{cite-2yzc-fsm3}
\end{enumerate}

\end{document}